\documentclass[11pt]{article}

\usepackage[margin=1in]{geometry}
\usepackage[T1]{fontenc}
\usepackage[utf8]{inputenc}
\usepackage{lmodern}
\usepackage{microtype}
\usepackage{amsmath,amssymb,amsthm}
\usepackage{booktabs}
\usepackage{tabularx}
\usepackage{array}
\usepackage[table]{xcolor}
\usepackage{hyperref}
\usepackage{enumitem}
\usepackage{graphicx}
\usepackage{float}
\usepackage{csquotes}

\usepackage{listings}

\lstdefinestyle{promptstyle}{
  basicstyle=\ttfamily\footnotesize,
  breaklines=true,
  breakatwhitespace=false,
  columns=fullflexible,
  keepspaces=true,
  inputencoding=utf8,
  extendedchars=true,
  literate={—}{{---}}1
           {–}{{--}}1
           {“}{{``}}1
           {”}{{''}}1
           {‘}{{`}}1
           {’}{{'}}1,
  frame=single,
  framerule=0.3pt,
  rulecolor=\color{black!35},
  backgroundcolor=\color{black!03},
  xleftmargin=0.5em,
  xrightmargin=0.5em,
  aboveskip=1em,
  belowskip=1em,
  numbers=left,
  numberstyle=\tiny\color{black!55},
  numbersep=0.7em,
  showstringspaces=false,
  tabsize=2
}

\hypersetup{
  colorlinks=true,
  linkcolor=blue,
  citecolor=blue,
  urlcolor=blue,
  pdftitle={The Obsolescence of High-Level Code: Agentic Generation from Semantic Artifacts},
  pdfauthor={Eduardo Di Santi},
  pdfsubject={Agentic software generation and semantic software artifacts},
  pdfkeywords={agentic software generation, semantic artifacts, software engineering, model-driven development, AI coding agents, verification},
  pdfcreator={LaTeX with hyperref},
  pdfproducer={LaTeX with hyperref}
}
 
\title{\textbf{The Obsolescence of High-Level Code:} \\ \vspace{0.5em}
{\large Agentic Generation from Semantic Artifacts}}
\author{Eduardo Di Santi\\[0.4em]
\normalsize College of Engineering and Applied Science\\
\normalsize University of Colorado Boulder\\[0.2em]
\small \texttt{eduardo.disanti@colorado.edu}}
\date{August 2026}

\begin{document}

\maketitle

\begin{abstract}
This paper argues that high-level source code need not remain the primary persistent artifact of software development when AI agents can generate, validate, and regenerate executable implementations from semantic intent. We place this claim in historical context by revisiting application generators, CASE tools, model-driven development, and platforms such as GeneXus, where developers already maintained higher-level representations from which code and other artifacts were derived.

We propose a unifying view in which generation is the fundamental operation: given a semantic specification $S=(R,C,I,T,O)$, maintained as the \emph{semantic authority}, an agent produces a generated executable artifact $A$. Functional requirements, contracts, and interfaces are assessed as semantic obligations, while tests and operational objectives are assessed by an explicit external oracle $V(A,S)$. We operationalize this framework through an experiment in independent generation: multiple LLVM artifacts are generated directly from one frozen semantic authority and evaluated by the same external oracle.

The experiment reports regeneration coherence on an evaluated domain, not a proof of absolute correctness. It provides initial evidence that independently generated executable artifacts can produce the same observable outcomes under a common semantic authority and external oracle. We conclude by discussing reproducibility, specification-centered evolution, and the implications of treating implementation representation as a deployment choice rather than the primary maintained artifact.
\end{abstract}

\section{Introduction}

The central thesis of this paper is simple: the human-facing representation of software and the persistent machine-facing representation of software need not coincide. Historically, high-level source code has served both as a medium for human reasoning and as the persistent artifact from which execution is derived. Under an agentic development regime, these roles may be separated.

In such a regime, the developer need not maintain the implementation artifact directly. Instead, the developer may maintain semantic intent: requirements, contracts, tests, interfaces, constraints, and optimization objectives. An agent then generates an executable artifact adapted to a particular target representation, such as C++, Rust, LLVM IR, WASM, or another machine-near form.

\paragraph{Origin: Technical Note, April 2026.}
This paper extends the Technical Note
\emph{Rethinking High-Level Code as the Primary Software Artifact},
released in April 2026 \cite{tn2026}. The Technical Note advanced the
following propositions:
\begin{itemize}
    \item The human-facing representation of software and its persistent
    machine-facing representation need not coincide.
    
    \item Under an agentic regime, developers may maintain requirements,
    invariants, tests, interfaces, and objectives, while an agent maintains
    an executable representation closer to the compiler or hardware.
    
    \item An initial experiment compared a CPython implementation with
    agent-generated LLVM IR for a toy kernel. Although the LLVM artifact
    ran approximately $74\times$ faster, the Technical Note explicitly
    stated that this result did not resolve the architectural hypothesis.
    
    \item The note concluded with open research questions, including:
    \emph{How should semantic alignment be verified across generation and
    specification-evolution cycles?}
\end{itemize}

\paragraph{The Central Question.}
If the developer never reads the generated code, why keep writing
high-level source code at all? This is a deliberately provocative question,
not a claim that source inspection is never useful. The narrower architectural
claim is that, when a generated implementation is not the authoritative
maintained artifact, its representation can be selected primarily for
execution and deployment requirements rather than for human maintenance.

In GeneXus\cite{genexus}, developers did not read the generated code---whether Cobol,
Java, C\#, or RPG. They maintained a formal semantic representation:
data dictionary, business rules, transactions, reports. The generator
was deterministic; correctness was \emph{by construction} from the
metamodel. The maintained artifact was the semantic model; the
generated code was a disposable compilation target.

Today, the semantic representation is no longer a fixed vendor
metamodel. It is you---your intent, your domain judgment---plus
structured specifications: requirements, contracts, interfaces, tests,
and optimization objectives. We denote this as $S = (R, C, I, T, O)$.
The agent generates \emph{from} this specification, not \emph{toward}
a predefined model.

This shift creates two fundamental problems:
\begin{enumerate}
    \item \textbf{Hallucinations and silent errors.} A stochastic generator can
    violate $S$ in ways no human sees, because no human reads the output.
    
    \item \textbf{Incomplete or ambiguous specification.} Your $S$ may not
    capture everything that matters: edge cases, emergent properties,
implicit invariants.
\end{enumerate}

\paragraph{Pragmatic Mitigations (Not a Formal Guarantee).}
Today we bridge this gap with a composite oracle:
\begin{itemize}
    \item Unit and integration tests---point validation of observed behavior.
    \item Verification and validation layers: executable contracts,
    property-based testing, differential testing across targets.
    \item GeneXus-style correctness-by-construction for what \emph{can}
    be formalized today: types, schemas, contracts, statically
    verifiable invariants.
\end{itemize}
These are pragmatic guards. They are not a formal guarantee of semantic
preservation. The research problem is precisely to move from this
patchwork to a principled verification framework.

\paragraph{Contributions.}
This paper makes six contributions:
\begin{enumerate}[label=\arabic*.]
    \item It situates agentic generation within the longer history of
    application generators, CASE environments, model-driven systems,
    and low-code platforms.
    \item It proposes a semantic specification schema
    \[
    S = (R, C, I, T, O)
    \]
    as the maintained human-facing artifact.
    \item It defines a \emph{specification-authoritative regeneration
    protocol}: a frozen, versioned semantic artifact is the sole authority
    for independently regenerated implementations; no prior implementation
    is an authority for a later one. Each run records provenance, is isolated
    from previous artifacts, and is evaluated against a common external
    oracle. This operationalizes, for agentic generation, the established
    spec-as-source and program-synthesis traditions \cite{gulwani2017}.
    \item It argues that \emph{generation} is the fundamental operation,
    while regeneration is generation under specification evolution.
    \item It outlines experiments to test whether implementation language
    behaves increasingly like a deployment parameter rather than a
    maintained human artifact.
    \item It frames semantic verification as an open research problem and
    proposes a pragmatic composite oracle as an interim approach.
\end{enumerate}

\section{Historical Lineage}

\subsection{Application Generators and CASE Tools}

Long before contemporary AI coding agents, software engineering already
explored forms of separation between maintained representations and
generated implementations. Fourth-generation languages, application
generators, and computer-aided software engineering (CASE) environments
allowed developers to define data models, business rules, user interfaces, and
reports from which executable systems could be produced. Early CASE work also
explored automatic generation of executable code and documentation from
structured requirements specifications \cite{case1990}.

In these environments, the developer often did not maintain the
generated code directly. Instead, the maintained artifact was a
higher-level description, whether textual, tabular, visual, or
model-based. Regeneration was achieved by modifying this
representation and re-running the generator.

\subsection{Model-Driven Development}

Model-driven development and related approaches made this separation
more explicit \cite{mda,brambilla2017}. A platform-independent model could be
transformed into platform-specific representations and then into code. The main
strength of these systems was not merely productivity; it was that the
transformation was constrained by a formal or semi-formal metamodel and a
designed transformation pipeline. Industrial studies show both the practical
value of model-driven engineering and the organizational conditions that shape
its adoption \cite{hutchinson2014}.

This distinction matters. Traditional model-driven systems derive much
of their strength from fixed structure. Their guarantees depend on the
narrowness and formality of the maintained representation.

\subsection{GeneXus as a Critical Antecedent}

GeneXus is especially relevant because it demonstrates that the
maintained representation can stand above source code for substantial,
production-oriented software systems \cite{genexus,gondaJodal2007}. Developers did
not merely use GeneXus to generate isolated functions. They maintained a knowledge
base or conceptual representation from which the platform could derive application
code, database evolution artifacts, documentation, and other implementation-level
outputs.

This historical precedent is important for two reasons. First, it
prevents the mistaken claim that AI agents uniquely invented the
separation between maintained intent and generated implementation.
Second, it sharpens the actual research question: what happens when
this separation is generalized beyond a fixed business-domain metamodel
and into an open-ended agentic setting?

\subsection{From Fixed Generators to Agentic Generation}

The shift introduced by LLM-based agents is not simply that code can
be generated from prompts. The more significant shift is that semantic
intent may no longer need to be encoded in a predefined formal model
owned by a vendor platform. Instead, intent may be expressed through a
mixture of natural language requirements, interfaces, tests, contracts,
and optimization preferences.

This creates new power, but also new uncertainty. We gain flexibility
across domains and target representations, but we lose the strong
structural guarantees that fixed generators derive from their
metamodels. Generative programming and software-product-line research
provide important precedents for deriving implementations from domain
models and configuration knowledge \cite{czarnecki2000}. Program synthesis
provides a complementary tradition in which programs are derived from formal
constraints or input--output behavior \cite{gulwani2017}. Contemporary LLM-based
code generation extends the range of executable artifacts that can be produced,
but its stochastic behavior makes an independent semantic authority and external
validation particularly important \cite{llmcodeSurvey}.

\subsection{Compilation-Oriented Representations}

The proposal does not treat LLVM as the contribution of this paper. LLVM is
used as a concrete machine-near target representation: a typed, low-level
compiler infrastructure designed to preserve program information across
analysis and transformation stages \cite{lattnerAdve2004,llvmLangRef}. The broader
architectural claim is representation-independent: when an executable artifact
is regenerated from maintained semantic authority and assessed by an external
oracle, its deployment representation may be chosen for operational rather than
human-maintenance objectives.

\begin{table}[H]
\centering
\caption{Historical comparison of maintained and generated software artifacts.}
\label{tab:historical-comparison}

\renewcommand{\arraystretch}{1.28}
\setlength{\tabcolsep}{3.2pt}
\footnotesize

\begin{tabularx}{\textwidth}{
  >{\raggedright\arraybackslash\bfseries}p{2.15cm}
  >{\raggedright\arraybackslash}X
  >{\raggedright\arraybackslash}X
  >{\raggedright\arraybackslash}X
  >{\raggedright\arraybackslash}X
}
\toprule
\textbf{Dimension} &
\textbf{CASE / 4GL} &
\textbf{MDD / DSL} &
\textbf{GeneXus / low-code} &
\textbf{Agentic generation} \\
\midrule

Maintained artifact &
Forms, schemas, rules &
Model or DSL &
Knowledge base or model &
Semantic specification $S$ \\

Generated artifact &
Application code &
Platform-specific code &
Application stack &
Machine-facing executable \\

Transformation &
Template-driven &
Rule-driven &
Platform generator &
Agentic, stochastic \\

Domain scope &
Narrow &
Structured domains &
Business applications &
Potentially general \\

Regeneration &
Supported &
Supported &
Supported &
From updated $S$ \\

Formal structure &
Medium &
High &
Medium--high &
Mixed / partial \\

Human inspection of generated code &
Often unnecessary &
Often unnecessary &
Often unnecessary &
Not required by the proposed architecture \\

Validation approach &
Tool-specific &
Constrained mapping or by construction &
Platform-specific validation &
External composite oracle \\

\bottomrule
\end{tabularx}
\end{table}

\section{Semantic Specification}

\subsection{Definition}

We define the maintained artifact as a semantic specification
\[
S = (R, C, I, T, O)
\]
where:
\begin{itemize}
    \item $R$ denotes requirements,
    \item $C$ denotes contracts and invariants,
    \item $I$ denotes interfaces and schemas,
    \item $T$ denotes tests,
    \item $O$ denotes optimization objectives and operational preferences.
\end{itemize}

This representation is not assumed to be fully formal. Some components
may be highly structured, while others may remain partly natural-language
based. The point is not that $S$ is perfect, but that it is the artifact
the human maintains.

\subsection{Interpretation}

The specification $S$ is intended to capture semantic intent at a level
closer to design and validation than to implementation syntax. A
developer can change $S$ by adding a new requirement, tightening a
contract, changing an interface, or altering an optimization objective.
The implementation artifact can then be generated again from the
updated specification.

The crucial shift is that correctness is no longer phrased as
equivalence to a human-maintained source code base. Instead, the
relevant question becomes whether a generated artifact satisfies the
maintained semantic specification.

\section{Generation as the Fundamental Operation}

\subsection{Base Case}

Let a semantic specification be
\[
S=(R,C,I,T,O),
\]
where $R$ denotes functional requirements, $C$ contracts and invariants,
$I$ interfaces and schemas, $T$ tests, and $O$ operational objectives. A
generator maps the specification to an executable artifact:
\[
G:S\mapsto A.
\]

The notation $A\models S$ is too coarse, because tests and operational
objectives are not, in general, logical propositions directly satisfied by an
artifact. We therefore distinguish semantic obligations from validation.
For requirements, contracts, and interfaces, the intended relation is
\[
A\models(R,C,I).
\]
For the complete specification, the relevant practical claim is instead
\[
\boxed{V(A,S)=\mathrm{PASS}},
\]
where $V$ is an external validation process. It may be decomposed as
\[
V(A,S)=V_{R,C,I}(A,S)\land V_T(A,T)\land V_O(A,O).
\]
This formulation makes explicit that tests provide evidence and that
operational objectives are measured against stated criteria.

\subsection{Specification-Centered Evolution}

Generation remains the same operation after a requirement change. If a
specification changes from $S_0$ to $S_1$, the intended workflow is
\[
A_0=G(S_0),\qquad A_1=G(S_1),
\]
rather than treating an implementation patch $A_0\rightarrow A_1$ as the
primary source of truth. The resulting artifacts may be structurally
unrelated while remaining conformant to their respective specifications.

This is an architectural consequence rather than a principal experiment in
this paper. Migration and compatibility mechanisms should be anchored to
explicit specification changes and their validation obligations, not solely to
implementation diffs.

This view does not require every executable artifact in a large system to be
regenerated after every localized change. It requires only that the maintained
representation identify the changed semantic intent and that the relevant
executable realization be regenerated and revalidated from that intent. How a
production system determines this scope is an engineering concern outside the
present argument. The conceptual point is narrower: even when regeneration is
selective, the authoritative change occurs in the semantic representation, not
in human-maintained implementation code.

\subsection{Why This Matters}

The distinction relocates the durable engineering asset from a particular
source representation to a specification and its validation evidence. It does
not eliminate engineering judgment: people must still formulate requirements,
choose constraints, assess evidence, and intervene when validation is
inadequate. It does, however, permit the implementation representation to be
chosen independently of its role as a human-maintained document.

\section{Experiment: Independent Generation from a Frozen Semantic Specification}

\subsection{Research Question and Design}

The experiment evaluates whether one frozen semantic specification $S_0$ can
be independently realized as multiple executable artifacts:
\[
S_0\longrightarrow A_1,A_2,\ldots,A_n.
\]
Each artifact is generated directly from $S_0$. Critically, no generated
implementation is used as input, context, or intermediate representation for
another generation:
\[
A_i\not\rightarrow A_{i+1}.
\]
For the LLVM experiment, Python is not an intermediate representation.

All artifacts are evaluated by a common external validation oracle:
\[
V(A_i,S_0)=\mathrm{PASS}.
\]
This design separates repeated semantic realization from iterative rewriting
of an earlier implementation.

\begin{figure}[htbp]
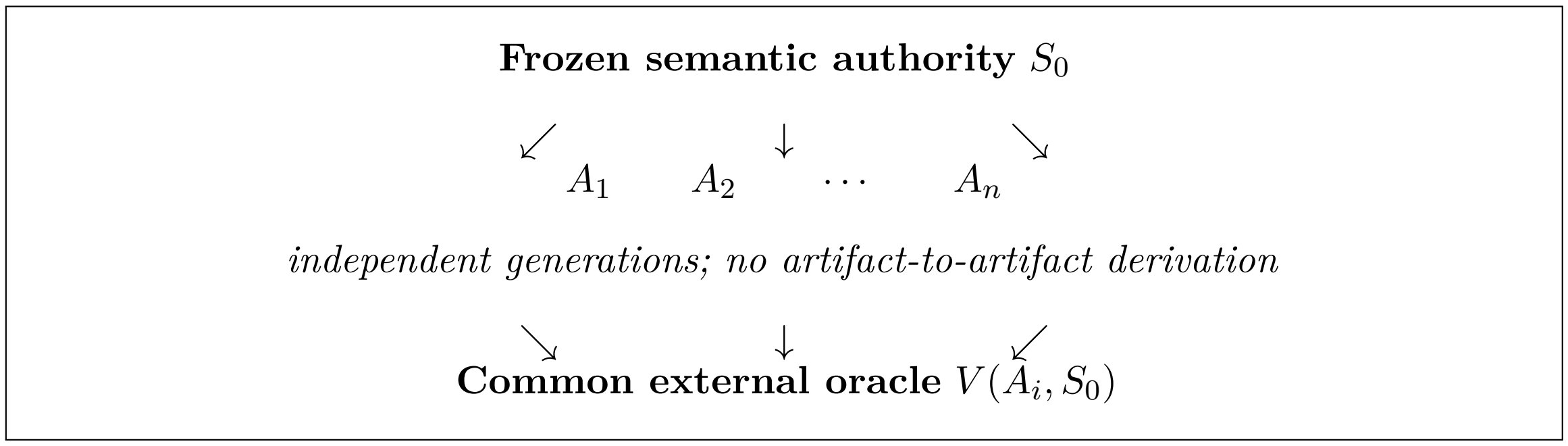

\centering
\setlength{\fboxsep}{10pt}
\fbox{\begin{minipage}{0.88\linewidth}
\centering
\textbf{Frozen semantic authority} $S_0$\\[0.8em]
$\swarrow\hspace{5.5em}\downarrow\hspace{5.5em}\searrow$\\[-0.2em]
$A_1\hspace{2.0em}A_2\hspace{2.0em}\cdots\hspace{2.0em}A_n$\\[0.8em]
\textit{independent generations; no artifact-to-artifact derivation}\\[0.8em]
$\searrow\hspace{5.5em}\downarrow\hspace{5.5em}\swarrow$\\[-0.2em]
\textbf{Common external oracle} $V(A_i,S_0)$
\end{minipage}}
\caption{Independent-generation protocol. Each executable artifact is generated directly from the same frozen semantic authority and evaluated by the same external oracle; the principal feasibility campaign fixes $n=10$ per domain.}
\label{fig:independent-generation-protocol}
\end{figure}

\begin{table}[htbp]
\centering
\caption{Experimental protocol and current realization.}
\label{tab:experimental-protocol}
\small
\begin{tabularx}{\linewidth}{@{}lX X@{}}
\toprule
\textbf{Element} & \textbf{\texttt{pricing\_demo}} & \textbf{\texttt{authorization\_demo}} \\
\midrule
Frozen semantic authority & Pricing requirements, monetary arithmetic, rounding, tax, and jurisdiction handling & Attribute-based authorization rules, strict JSON interface, and deny-overrides-allow policy \\
Generated artifacts & $n$ independent LLVM artifacts; principal campaign $n=10$ & $n$ independent LLVM artifacts; principal campaign $n=10$ \\
Observable interface & JSON pricing result or explicit error on stdout & JSON decision or explicit error on stdout, plus process exit status \\
External oracle & Golden, property-based, and differential tests; 1,000 randomized cases & Golden and invalid-input cases, rule boundaries, and 1,000 seeded randomized cases \\
Coherence criterion & All evaluated artifacts pass the common oracle for $S_0$ & All evaluated artifacts pass the common oracle for $S_0$ \\
\bottomrule
\end{tabularx}
\end{table}

\subsection{Experimental Realization}

For each domain $d$, the study evaluates $n$ independently generated LLVM
artifacts derived directly from one frozen semantic specification $S_d$:
\[
S_d \longrightarrow A_{d,1}, A_{d,2}, \ldots, A_{d,n}.
\]
No generated artifact is used as input, context, or intermediate
representation for another generation. All artifacts are evaluated against the
same domain-specific external oracle $V_d$.

The principal feasibility campaign fixes $n=10$ independent regenerations per
domain: ten pricing artifacts and ten authorization artifacts, for a total of
twenty independently generated executable artifacts. Each run retains the
frozen specification, prompt record, generated candidate, provenance metadata,
and validation transcript. Results are reported only after each campaign is
complete. They establish pass status under the stated specification, test suite,
and operational checks; they do not establish absolute correctness or formal
semantic preservation beyond those checks.

\subsection{The \texttt{pricing\_demo} Experiment}

The \texttt{pricing\_demo} experiment instantiates the proposed architecture
with a small but semantically nontrivial order-pricing engine. The frozen
semantic specification $S_0$ defines requirements for subtotal computation,
tiered volume discounts, jurisdiction-dependent sales tax, banker's rounding,
integer monetary arithmetic, and fail-closed handling of unknown
jurisdictions.

The frozen specification $S_0$, reproduced in
Appendix~\ref{app:specification}, was the sole authoritative source of
semantic requirements for every generated artifact. No implementation artifact
was authoritative for a later generation.

The specification also defines explicit contracts. In particular, totals must
be nonnegative, discount must remain between zero and the subtotal, tax must
be nonnegative, and the accounting identity
\[\mathrm{subtotal} - \mathrm{discount} + \mathrm{tax} = \mathrm{total}\]
must hold exactly. The common interface accepts a JSON order on standard input
and returns either a JSON pricing result or an explicit error for an unknown
jurisdiction.

An initial Python implementation was generated with the Terra Thinking model.
This implementation served as an early executable realization of the semantic
specification, but it was not used as an intermediate representation for the
subsequent LLVM generations. The principal pricing campaign generates ten LLVM
IR artifacts, denoted $A_{\mathrm{price},1}, \ldots,
A_{\mathrm{price},10}$, independently from the same frozen specification
$S_0$.

Thus, the experimental lineage is
\[S_0 \longrightarrow A_{\mathrm{Python}}\]
for the initial Python realization, and independently
\[S_0 \longrightarrow A_{\mathrm{price},1}, \ldots, A_{\mathrm{price},10}.\]
No LLVM artifact was generated by transforming, translating, or iteratively
repairing another generated implementation. In particular, Python was not an
intermediate representation in the LLVM-generation procedure.

\subsection{Authorization Regeneration-Coherence Experiment}

The \texttt{authorization\_demo} provides a second, deliberately bounded
case study. Its frozen specification defines an attribute-based authorization
engine with a fixed policy table, strict JSON acceptance rules, default denial,
and deny-overrides-allow semantics. The purpose of this case study is not to
claim the correctness of one generated program, nor to evaluate a language
model. It tests the architectural question of whether independent
regenerations from the same semantic specification remain observationally
coherent at the executable interface.

The principal authorization campaign generates ten LLVM IR artifacts, denoted
$A_{\mathrm{auth},1}, \ldots, A_{\mathrm{auth},10}$, independently from the
frozen authorization specification, reproduced in
Appendix~\ref{app:authorization-specification}. The independent LLVM
generation prompt is reproduced in
Appendix~\ref{app:authorization-prompts}; during each generation, its
specification placeholder was replaced with the exact versioned text shown in
Listing~\ref{lst:authorization-specification}. The validation harness compiled each artifact for
the stated Apple Silicon target and compared its stdout JSON response and
process exit status with an independently written reference oracle. The suite
covered the specified golden and boundary cases, malformed and
unknown-identifier requests, deny-overrides-allow behavior including the
intentional P8/P9 conflict, default denial, and 1,000 seeded randomized
requests containing valid and invalid inputs.

For each authorization artifact, the outcome is recorded as
$V_{\mathrm{auth}}(A_{\mathrm{auth},i},S_{\mathrm{auth}}) \in
\{\mathrm{PASS},\mathrm{FAIL}\}$, for $1 \leq i \leq 10$. The completed
campaign reports the number of passing artifacts and its observed stability;
all ten artifacts successfully passed the external oracle. This provides evidence of
regeneration coherence under the stated evaluation procedure, not a proof of
universal semantic equivalence or absolute program correctness. The value of the result is that the maintained authority can
remain the semantic specification and its oracle while the executable artifact
is regenerated in a machine-near representation selected for deployment
criteria such as latency, memory use, energy use, target compatibility, and
optimization opportunity.

\subsection{Generation Protocol and Prompt Record}

The experiment used a documented generation protocol rather than an
undocumented sequence of conversational interactions. The protocol separated
four roles: (i) agent framing, (ii) generation of an executable reference
implementation, (iii) generation of validation artifacts, and (iv) independent
generation of LLVM IR artifacts from the frozen semantic specification.

The initial Python realization was generated with the Terra Thinking model.
The subsequent LLVM artifacts were generated in separate runs from $S_0$ using
GPT-5.6 Luna and the LLVM-generation prompt specified in Appendix~\ref{app:prompts}. The
generation prompt required a JSON-compatible executable implementation of the
pricing specification and explicitly prohibited the use of Python or another
generated implementation as an intermediate representation.

The validation artifacts were specified separately from the implementation
artifacts. The protocol includes prompts for golden test cases, a
property-based test harness, a differential-test harness, and a composite
oracle runner. This separation is important: the validation procedure must not
be derived from the particular structure of a generated LLVM artifact.

For reproducibility, Appendix~\ref{app:prompts} reproduces the complete prompt
record used in the experiment. The corresponding machine-readable prompt files,
semantic specification, generated artifacts, validation harnesses, compilation
commands, and execution results are retained in the accompanying repository.

\subsection{Results}

Each LLVM artifact was compiled and evaluated against the same deterministic
external oracle derived from $S_0$. The validation procedure included nine
golden and boundary cases, four unknown-jurisdiction cases, and 1,000
randomly generated cases. The golden and boundary set covers the empty order,
small orders, discount-tier boundaries, representative multi-line orders, and
rounding-sensitive examples. Unknown jurisdictions are required to produce an
explicit error rather than silently defaulting to a zero tax rate.

\begin{table}[H]
\centering
\caption{Results of the $n=10$ pricing feasibility campaign.}
\label{tab:pricing-demo-results}
\begin{tabular}{lcccc}
\toprule
Artifact & Golden/boundary & Unknown jurisdiction & Random & Overall \\
\midrule
$A_{\mathrm{price},1}$  & 9/9 & 4/4 & 1000/1000 & PASS \\
$A_{\mathrm{price},2}$  & 9/9 & 4/4 & 1000/1000 & PASS \\
$A_{\mathrm{price},3}$  & 9/9 & 4/4 & 1000/1000 & PASS \\
$A_{\mathrm{price},4}$  & 9/9 & 4/4 & 1000/1000 & PASS \\
$A_{\mathrm{price},5}$  & 9/9 & 4/4 & 1000/1000 & PASS \\
$A_{\mathrm{price},6}$  & 9/9 & 4/4 & 1000/1000 & PASS \\
$A_{\mathrm{price},7}$  & 9/9 & 4/4 & 1000/1000 & PASS \\
$A_{\mathrm{price},8}$  & 9/9 & 4/4 & 1000/1000 & PASS \\
$A_{\mathrm{price},9}$  & 9/9 & 4/4 & 1000/1000 & PASS \\
$A_{\mathrm{price},10}$ & 9/9 & 4/4 & 1000/1000 & PASS \\
\bottomrule
\end{tabular}
\end{table}

Each pricing artifact is evaluated on 1,013 cases: nine golden and boundary cases, four unknown-jurisdiction cases, and 1,000 randomly generated cases. At $n=10$, a completed pricing campaign therefore comprises 10,130 validation executions under its common oracle.

\begin{table}[htbp]
\centering
\caption{Results of the $n=10$ authorization feasibility campaign.}
\label{tab:auth-demo-results}
\begin{tabular}{lcccc}
\toprule
Artifact & Golden/boundary & Invalid/unknown input & Random & Overall \\
\midrule
$A_{\mathrm{auth},1}$  & 12/12 & 8/8 & 1000/1000 & PASS \\
$A_{\mathrm{auth},2}$  & 12/12 & 8/8 & 1000/1000 & PASS \\
$A_{\mathrm{auth},3}$  & 12/12 & 8/8 & 1000/1000 & PASS \\
$A_{\mathrm{auth},4}$  & 12/12 & 8/8 & 1000/1000 & PASS \\
$A_{\mathrm{auth},5}$  & 12/12 & 8/8 & 1000/1000 & PASS \\
$A_{\mathrm{auth},6}$  & 12/12 & 8/8 & 1000/1000 & PASS \\
$A_{\mathrm{auth},7}$  & 12/12 & 8/8 & 1000/1000 & PASS \\
$A_{\mathrm{auth},8}$  & 12/12 & 8/8 & 1000/1000 & PASS \\
$A_{\mathrm{auth},9}$  & 12/12 & 8/8 & 1000/1000 & PASS \\
$A_{\mathrm{auth},10}$ & 12/12 & 8/8 & 1000/1000 & PASS \\
\bottomrule
\end{tabular}
\end{table}

The corresponding authorization campaign reports its ten artifact outcomes under its domain-specific oracle, for a total of 10,200 validation executions. All ten artifacts were generated by the same stochastic model (GPT-5.6 Luna, operating at temperature $>0$) to evaluate generation stability across independent runs.

\subsection{Generation Stability}

For a frozen specification $S_d$, a fixed generation protocol, and $n$
independent generations, observed generation stability is defined as
\[
\mathrm{Stability}(S_d,n)
= \frac{1}{n} \sum_{i=1}^{n}
\mathbf{1}\left[
V_d(A_{d,i},S_d)=\mathrm{PASS}
\right].
\]
Here $\mathbf{1}[\cdot]$ equals one when the stated condition holds and zero
otherwise. For the principal feasibility campaign, $n=10$; hence
\begin{equation}
\mathrm{Stability}(S_d, 10) = \frac{\#\{A_{d,i} \mid V_d(A_{d,i}, S_d) = \mathrm{PASS}\}}{10}.
\end{equation}
For the completed campaigns, $\mathrm{Stability}(S_{\mathrm{price}}, 10) = 10/10 = 1.00$ and $\mathrm{Stability}(S_{\mathrm{auth}}, 10) = 10/10 = 1.00$.

Stability is reported separately for pricing and authorization, rather than as
a single pooled primary metric, because their specifications and oracles differ.
It is an observed measure for the stated specification, prompt, execution
environment, and validation oracle, not an estimate of universal model
reliability or a proof of semantic equivalence beyond the evaluated oracle.

\subsection{Implementation Representation as Deployment Choice}

If distinct artifacts generated from a common semantic specification pass a
common oracle, implementation representation can vary without changing the
maintained semantic intent. This motivates treating language or target form as
a deployment parameter, constrained by performance, portability, safety, and
runtime requirements. It does not imply that all representations are equally
suitable, nor that human review is unnecessary in every assurance context.

\section{What the Experiment Establishes and Does Not Establish}

The empirical claim of this paper is deliberately narrow. The experiment does
not evaluate a language model, rank code-generation systems, or establish the
absolute correctness of any one generated program. It evaluates whether
independently generated, machine-near executable artifacts can remain
observationally coherent with one frozen semantic authority under one external
oracle on a defined request and input domain.

The two case studies provide initial evidence for that claim. For each frozen
specification, independently generated LLVM artifacts compiled for the target
platform and produced the expected observable results under the applicable
oracle: stdout JSON response, process exit status, and the tested functional
outcomes. In the authorization case, this includes strict rejection behavior,
default denial, and deny-overrides-allow; in the pricing case, it includes the
specified arithmetic and policy outcomes. The evidence is therefore about
regeneration coherence, not about an individual artifact being privileged as
the correct implementation.

Three distinct levels of claim should remain separate. First, the
\emph{architectural thesis} is that high-level source code need not be the
primary persistent artifact when semantic authority and external validation
are maintained. Second, the \emph{experimental evidence} is that the evaluated
LLVM regenerations produced coherent observable behavior for the stated cases.
Third, the \emph{operational hypothesis} is that a machine-near deployment
artifact may be selected or regenerated for latency, binary size, memory,
energy use, target compatibility, or optimization opportunity. This experiment
does not yet quantify those operational properties.

The study also does not prove universal semantic equivalence, demonstrate
coherence outside the evaluated domains, or establish that all future
regenerations will pass. Its contribution is a reproducible protocol and an
initial empirical result: the semantic authority, rather than a prior source
implementation, can remain the maintained reference while independently
generated executable artifacts are assessed against a common external oracle.

\section{Validation Without Human-Maintained Source}

\subsection{Validation Claim}

When generated code is not the authoritative maintained artifact, confidence
must be attached to the specification, the generator, and an explicit
validation procedure. The appropriate claim is not that a generated artifact
is correct in an absolute sense, but that it passes a named and reproducible
oracle for a given semantic specification:
\[
V(A,S)=\mathrm{PASS}.
\]

\subsection{The Asymmetric Oracle Principle}

A fundamental vulnerability in early agentic systems is the \emph{Oracle Tautology}: if an LLM generates the artifact $A$ and the same (or similar) LLM is used to evaluate its correctness ($V_{LLM}(G(S)) \to \mathrm{PASS}$), the model is likely to hallucinate the validation, grading its own flawed logic as correct \cite{bubeck2023sparks}. 

To prevent this, validation must adhere to the \textbf{Asymmetric Oracle Principle}: the evaluation mechanism $V$ must be computationally asymmetrical to the generator $G$. It must be grounded in absolute, deterministic execution rather than stochastic review. 

\subsection{Composite Validation Oracles}

A practical, asymmetric oracle may combine contracts, example-based tests, property-based checks,
differential testing, mutation testing, interface validation, and measurement of operational objectives.
Furthermore, the oracle is not restricted to traditional software tests: it may be a physical emulator, a Programmable Logic Controller (PLC), or a separate bounded mathematical model. Crucially, maintaining an oracle requires managing a single, independently verifiable artifact rather than a complete, interlocking system architecture.

These mechanisms test different failure modes and may provide useful cumulative evidence. They are
not a formal guarantee of semantic preservation, and passing one oracle does
not establish conformance to unstated requirements.

The current experiment uses a shared external procedure containing golden and
boundary cases, unknown-jurisdiction cases, and randomized cases. Its value is
that each independently generated artifact is assessed against the same
criterion rather than against another generated implementation.

\subsection{Reproducibility}

A reproducible study must preserve the frozen semantic specification, the
artifact-generation protocol, prompts and model or tool versions where
applicable (all 20 reported artifacts in this study were generated by GPT-5.6 Luna), compilation settings, validation-oracle implementation, random
seeds, hardware and runtime environment, generated artifacts or their hashes,
and complete pass/fail results. These materials permit later researchers to
separate changes in the specification, generator, environment, and oracle.

\section{Discussion}

\subsection{From Technical Debt to Specification Debt}

A direct consequence of this architecture is the obsolescence of traditional technical debt \cite{cunningham1992wycash}. Poor
abstractions, spaghetti code, and missing comments become irrelevant when the artifact is exclusively
machine-generated and machine-read. However, this is replaced by a new burden: \textbf{Specification
Debt}. This arises when human intent contradicts the formal contracts, when the specification relies
on obsolete assumptions, or when the external oracle $V$ is too weak to catch generator hallucinations or ``reward hacking,'' where the model produces valid but unintended behavior merely to pass incomplete tests. This vulnerability exactly mirrors human developers writing brittle code to satisfy a flawed test suite. The architectural claim is not that agentic generation immediately achieves superhuman reliability, but rather that it achieves comparable operational coherence at radically different speeds and costs. Refactoring in the future will mean refactoring specifications and oracles, not source code.

\subsection{The Role of Human-Maintained Code}

The strongest version of the argument is not that source code
disappears. In some domains, human-maintained code will remain useful,
necessary, or legally required. The stronger and more interesting
claim is narrower: source code may cease to be the universally primary
software artifact.

GeneXus and related systems already showed that this could happen
within structured domains. The developer maintained a semantic model;
the generated code was never read. Correctness came from the
deterministic generator and the fixed metamodel.

Agentic generation raises the possibility that the same separation can
be generalized across much wider classes of software, provided that
semantic intent can be maintained and validation can be made
sufficiently robust. The central challenge is no longer generation per
se---it is validation when the generator is stochastic and the
specification is open-ended.

This also changes the role of the developer. The developer becomes
less responsible for maintaining every implementation detail and more
responsible for maintaining the semantic contract under which
implementations are generated, selected, and validated.

\section{Limitations and Future Work}

This paper reports an initial empirical feasibility campaign rather than
a complete demonstration of the architecture's generality. The principal
design evaluates two bounded domains, \texttt{pricing\_demo} and
\texttt{authorization\_demo}, with $n=10$ independently generated LLVM
artifacts per frozen semantic specification. Results are reported separately
by domain with complete generation and validation metadata.

Both campaigns were executed to completion; all 20 generated artifacts (10 pricing, 10 authorization) passed their respective oracles with zero failures.

Several limitations remain. First, zero observed failures at $n=10$ bounds the failure rate loosely; a larger campaign is needed to estimate stability with tighter confidence. Second, the current result concerns ``Zero-State Generation,'' where purely functional logic is regenerated from scratch. Most enterprise software, however, relies on persistent relational data. If an agent discards a previous artifact $A_0$ to generate a new artifact $A_1$, it risks \textbf{Regeneration Amnesia}---the loss of tacit structural decisions that preserve existing data states. Solving this via \textbf{Cumulative Regeneration} is the next critical frontier. As demonstrated historically by GeneXus's database reorganization features, the initial generation from $S_0$ must produce both the executable application and the physical database schema. Upon specification evolution to $S_1$, the generator must read the existing physical schema and output not only the new executable $A_1$, but also the necessary conversion scripts (e.g., \texttt{ALTER TABLE} commands and data-migration logic to compute and persist new values). Validating this stateful reorganization under stochastic generation remains a major open challenge.

Third, validation success under the stated oracle
is evidence of semantic conformity, not a proof of absolute correctness.
Fourth, the current study evaluates isolated, single-function components. As systems scale, semantic specifications must describe networks of interoperating modules. Generating artifacts that compose robustly across static interfaces without forcing the generation of a massive, monolithic executable is a critical architectural requirement for future research. Fifth, the current study does not yet report a systematic evaluation of
operational objectives such as latency, memory behavior, or compiler-level
robustness across generations.

Future work should therefore proceed along four directions:
\begin{enumerate}[label=\arabic*.]
    \item extend the independent-generation experiment to at least ten LLVM
    artifacts and report generation stability with complete generation and
    validation metadata;
    \item replicate the experiment across additional target representations,
    such as Rust, C++, WASM, or other deployment-oriented forms;
    \item evaluate specification evolution through controlled changes from
    $S_0$ to subsequent versions, regenerating artifacts from each revised
    specification rather than patching prior implementations;
    \item strengthen the validation oracle with property-based testing,
    differential testing across targets, mutation testing, operational
    measurements, and, where feasible, formal methods.
\end{enumerate}

A subsequent paper may focus specifically on verification under stochastic
generation. The contribution of the present paper is narrower: it establishes
the conceptual and historical framework and provides an initial experimental
demonstration that independently generated machine-facing artifacts can pass
a common validation procedure while the semantic specification remains the
maintained authoritative artifact.

\bibliographystyle{plain}

\appendix

\section{Frozen Semantic Specification for \texttt{pricing\_demo}}
\label{app:specification}

The frozen semantic specification $S_0$ is the authoritative maintained
artifact for the \texttt{pricing\_demo} experiment. It defines the functional
requirements, contracts and invariants, interface, validation obligations, and
operational objectives from which executable artifacts were independently
generated.

The specification below is reproduced directly from the versioned experiment
record. It is included before the prompt record because each generation and
validation instruction is interpreted relative to this specification.

\begin{lstlisting}[
  style=promptstyle,
  caption={Frozen semantic specification $S_0$ for the order-pricing engine.},
  label={lst:pricing-specification}
]
# Semantic Specification S — Order Pricing Engine (v0)

S = (R, C, I, T, O)

## R — Requirements
1. Given a list of line items (sku, unit_price, quantity), compute the subtotal.
2. Apply a tiered volume discount based on the subtotal (before tax), using a
   fixed, ordered table of (threshold, rate) pairs. The highest threshold met
   determines the rate; discount = round(subtotal * rate).
3. Apply jurisdiction-based sales tax on the discounted subtotal:
   tax = round((subtotal - discount) * tax_rate[jurisdiction]).
4. Reject unknown jurisdictions (fail closed, not silently default to 0% tax).
5. All monetary values are integers in the currency's minor unit (cents for
   USD/EUR, whole yen for JPY, thousandths of dinar for BHD) to avoid binary
   floating-point rounding errors.
6. Rounding is banker's rounding (round-half-to-even) at every intermediate
   step (discount, tax), matching common accounting practice.

## C — Contracts / Invariants
- C1: total >= 0
- C2: 0 <= discount <= subtotal
- C3: tax >= 0
- C4: subtotal - discount + tax == total   (exact equality; integers, no epsilon)
- C5: given the same input, output is deterministic (no hidden randomness or
      wall-clock dependence)
- C6: unknown jurisdiction => explicit error, never a silent default
- C7: empty item list => subtotal = discount = tax = total = 0 (no error)

## I — Interface
```
price_order(items: List[LineItem], jurisdiction: str) -> PricingResult

LineItem   = { sku: str, unit_price_minor: int (>=0), quantity: int (>=1) }
PricingResult = { subtotal: int, discount: int, tax: int, total: int }
```
Wire format: JSON on stdin -> JSON on stdout (used identically by every
generated target, enabling differential testing across languages).

## T — Tests
- Golden cases (tests/golden_cases.json): hand-verified subtotal/discount/tax/
  total for representative orders (empty cart, single item, each discount
  tier boundary, unknown jurisdiction).
- Property-based tests (tests/property_test.py, Hypothesis): C1-C4 checked
  against thousands of randomly generated orders.
- Differential tests (tests/differential_test.py): the same random orders are
  run through every generated target (Python now; Rust, C++, WASM in later
  iterations) and outputs must match exactly.

## O — Optimization Objectives / Operational Preferences
- O1: p99 latency < 1ms per order on reference hardware for the checkout path.
- O2: no floating-point arithmetic anywhere in the money path.
- O3: memory allocation independent of jurisdiction table size (O(1) lookup).

---

## Discount table (v0 — fixed for S0, evolves in later specification versions)

| Subtotal >= (minor units) | Rate |
|---|---|
| 0        | 0%  |
| 100000   | 5%  |
| 500000   | 10% |
| 1000000  | 15% |

## Jurisdiction tax table (v0)

| Jurisdiction | Rate  |
|---|---|
| US-CA | 7.25% |
| US-NY | 8.00% |
| US-OR | 0.00% |
| EU-DE | 19.00% |
| EU-IE | 23.00% |
\end{lstlisting}

\section{Complete Prompt Record for \texttt{pricing\_demo}}
\label{app:prompts}

This appendix reproduces the generation and validation prompts used in the
\texttt{pricing\_demo} experiment. The prompts are included after $S_0$
because the specification is their shared semantic input. Files that contain a
placeholder such as \texttt{<PASTE spec/S.md HERE>} are prompt templates; in
the executed protocol, that placeholder was replaced with the exact frozen
specification reproduced in Listing~\ref{lst:pricing-specification}.

\subsection{Agent Framing}
\label{app:prompt-agent-framing}

\begin{lstlisting}[
  style=promptstyle,
  caption={Agent framing prompt.},
  label={lst:agent-framing}
]
You are a code-generation agent. You will be given a semantic specification
S = (R, C, I, T, O) for a software component. Your task is to generate an
EXECUTABLE ARTIFACT that satisfies S. You must not invent requirements not
present in S. You must not weaken any contract in C. If S is ambiguous on a
point, state your assumption explicitly in a comment rather than silently
guessing. Do not add features beyond R. Optimize only for what O specifies.
\end{lstlisting}

\subsection{Python Reference Implementation}

\begin{lstlisting}[
  style=promptstyle,
  caption={Prompt used to generate the initial Python implementation.},
  label={lst:python-generation}
]
<system framing from Prompt 0>

Here is the specification (verbatim contents of spec/S.md):

<PASTE spec/S.md HERE>

Generate a Python 3 module implementing price_order(items, jurisdiction)
exactly as described in I. Enforce every contract in C1-C7 with runtime
assertions on the result before returning it. Use only integer arithmetic
for money (O2) -- no floats anywhere in the calculation path. Implement
banker's rounding (round-half-to-even) using integer division and modulo,
not Python's built-in round() (which has float-related edge cases).
Raise a dedicated exception type for unknown jurisdictions (C6) rather than
returning a sentinel value. Include a __main__ block that reads a JSON
payload {"items": [...], "jurisdiction": "..."} from stdin and writes
{"subtotal": ..., "discount": ..., "tax": ..., "total": ...} to stdout, or
{"error": "..."} with a non-zero exit code on UnknownJurisdictionError.
\end{lstlisting}

\subsection{LLVM Implementation Generation}

\begin{lstlisting}[
  style=promptstyle,
  caption={Prompt used for each independent LLVM generation.},
  label={lst:llvm-generation}
]
You are a code-generation agent.

You will be given a semantic specification S = (R, C, I, T, O) for a software component.

Your task is to generate a COMPLETE, EXECUTABLE LLVM IR implementation of the component described by S.

CRITICAL PRINCIPLE

The LLVM implementation must be generated DIRECTLY from S.

Do NOT translate, inspect, or derive the implementation from any Python, C, C++, Rust, or other high-level implementation.

The semantic specification S is the sole authoritative source of requirements.

You must not invent requirements that are not present in S.
You must not weaken, reinterpret, or silently relax any contract in C.
You must not add functionality beyond R.
You must optimize only according to O.

If S is ambiguous or underspecified, state the assumption explicitly before the LLVM IR. Do not silently guess.

TARGET

Generate LLVM IR (.ll) suitable for compilation with clang/LLVM tooling available in the execution environment.

The generated artifact must implement the interface specified in I.

The generated artifact must expose the same externally observable behavior defined by S, independently of any other implementation.

Do not generate LLVM IR that depends on another implementation.

NUMERIC SEMANTICS

Follow the numeric semantics specified in S exactly.

In particular:

- Use integer arithmetic where S requires integer arithmetic.
- Do not introduce floating-point arithmetic where S specifies integer arithmetic.
- Implement rounding exactly as specified by S.
- Preserve the specified overflow behavior.
- Do not rely on undefined behavior for any operation whose behavior is constrained by S.
- Make integer widths explicit and appropriate to the constraints in S.

CONTRACTS

Every contract C1...Cn in S must be enforced or satisfied by the generated implementation.

Where a contract is an output invariant, ensure that the returned result satisfies it.

Where a contract specifies error or fail-closed behavior, implement that behavior explicitly.

Do not weaken a contract merely because enforcing it is inconvenient at LLVM level.

INTERFACE

Implement the exact interface described in I.

If S defines a JSON wire protocol, preserve the exact externally observable JSON schema and semantics.

Do not add fields.
Do not remove fields.
Do not change field names.
Do not change numeric semantics.
Do not change error semantics.

If a command-line interface is required by I, the generated artifact must be executable through that interface.

PORTABILITY AND LLVM

Use standard LLVM IR constructs.

Do not rely on target-specific instructions unless explicitly required by O or I.

Prefer explicit control flow and integer operations over opaque or implementation-dependent behavior.

Do not embed a second programming language inside the LLVM IR.

Do not generate pseudocode.

The output must be valid LLVM IR that can actually be assembled and executed.

VALIDATION REQUIREMENTS

The generated artifact must be suitable for validation against the tests defined by S.

The implementation must therefore expose deterministic behavior for identical inputs.

Do not introduce randomness unless explicitly required by S.

Do not introduce external state unless explicitly required by S.

Do not modify the specification or its tests.

ASSUMPTIONS

Before the LLVM IR, provide a short section:

ASSUMPTIONS

List only assumptions that are necessary because S is ambiguous or because the execution environment requires a choice not specified by S.

For each assumption, explain which part of S does not determine the choice.

Do not use assumptions to add requirements or functionality.

OUTPUT FORMAT

Produce exactly two logical sections:

1. ASSUMPTIONS

2. LLVM IR

The LLVM IR section must contain the complete contents of the .ll file in a single fenced code block.

Do not provide Python, Rust, C, C++, pseudocode, or a prose implementation instead of LLVM IR.

The implementation must be complete and self-contained except for explicitly documented external interfaces required by I.

FINAL SELF-CHECK

Before producing the answer, internally check that:

1. Every requirement in R is implemented.
2. Every contract in C is preserved.
3. Every interface requirement in I is respected.
4. Every testable behavior in T is representable by the implementation.
5. Every optimization constraint in O is respected.
6. No requirement was invented.
7. No contract was weakened.
8. No high-level implementation was used as an intermediate representation.
9. The output is actual LLVM IR, not pseudocode.
10. The generated artifact can be compiled by standard LLVM tooling.
\end{lstlisting}

\subsection{Golden-Test Generation}

\begin{lstlisting}[
  style=promptstyle,
  caption={Prompt for hand-verifiable golden and boundary test cases.},
  label={lst:golden-generation}
]
<system framing from Prompt 0>

Here is the specification (verbatim contents of spec/S.md):

<PASTE spec/S.md HERE>

Generate a JSON file of hand-verifiable golden test cases for price_order,
covering: (a) empty cart, (b) a single item below every discount
threshold, (c) an order exactly at each discount tier boundary
(off-by-zero) and one minor unit below it (off-by-one), (d) a
multi-jurisdiction spread including at least one zero-tax jurisdiction and
at least one high-tax jurisdiction, (e) an unknown-jurisdiction rejection
case, and (f) a large-quantity case that stresses integer overflow
boundaries relevant to O2. For every case with an expected numeric result,
show your arithmetic reasoning so the result can be independently
recomputed and verified -- do not simply assert a plausible-looking number.
\end{lstlisting}

\subsection{Property-Based Test Harness}

\begin{lstlisting}[
  style=promptstyle,
  caption={Prompt for the property-based validation harness.},
  label={lst:property-harness}
]
<system framing from Prompt 0>

Here is the specification (verbatim contents of spec/S.md):

<PASTE spec/S.md HERE>

Generate a property-based test script that randomly generates thousands of
orders (varying item count, unit price, quantity, and jurisdiction) and
checks that contracts C1 through C7 hold on every generated order, without
relying on any single hand-picked example. Also verify the fail-closed
behavior of C6 against several malformed or unknown jurisdiction strings
(including a case-sensitivity check and a trailing-whitespace check, since
S does not explicitly state whether jurisdiction matching is case- or
whitespace-sensitive -- surface this as an assumption). If the standard
property-testing library for this language is unavailable in the execution
environment, use a seeded pseudo-random generator instead and state this
substitution explicitly; the fixed seed must make the test run
reproducible (C5-consistent methodology).
\end{lstlisting}

\subsection{Differential-Test Harness}

\begin{lstlisting}[
  style=promptstyle,
  caption={Prompt for the differential-test harness.},
  label={lst:differential-harness}
]
<system framing from Prompt 0>

Here is the specification (verbatim contents of spec/S.md):

<PASTE spec/S.md HERE>

Two independent executable artifacts have been generated from this same
specification: a Python module and a Rust binary, both communicating via
the JSON wire format described in I. Generate a differential test script
that produces randomized orders, runs each order through BOTH artifacts
(the Python one in-process, the Rust one via subprocess using its stdin/
stdout JSON contract), and asserts the two outputs are byte-for-byte
identical. The script must report the count of trials, list every mismatch
with full input detail, and exit non-zero if any mismatch occurs. Use a
fixed random seed for reproducibility.
\end{lstlisting}

\subsection{Composite Oracle Runner}

\begin{lstlisting}[
  style=promptstyle,
  caption={Prompt for the composite validation-oracle runner.},
  label={lst:oracle-runner}
]
<system framing from Prompt 0>

Given that golden_test.py, property_test.py, and differential_test.py all
exist under tests/, and a Rust crate exists under rust/, generate a single
shell script that (1) builds the Rust release binary, (2) runs the golden
tests, (3) runs the property-based tests, (4) runs the differential tests,
in that order, stopping and returning non-zero on the first failing layer,
and printing a clear pass/fail summary for each layer.
\end{lstlisting}

\section{Frozen Semantic Specification for \texttt{authorization\_demo}}
\label{app:authorization-specification}

The frozen semantic specification $S_{\mathrm{auth}}$ is the maintained authoritative
artifact for the \texttt{authorization\_demo} regeneration-coherence
experiment. It defines the authorization requirements, invariants, JSON
interface, fixed policy table, validation obligations, and operational
objectives used for all ten independent LLVM generations.

The specification is reproduced directly from the versioned experiment record.
It appears before the generation prompt because that prompt is interpreted
relative to the exact specification shown in
Listing~\ref{lst:authorization-specification}.

\begin{lstlisting}[
  style=promptstyle,
  caption={Frozen semantic specification $S_{\mathrm{auth}}$ for the attribute-based authorization engine.},
  label={lst:authorization-specification}
]
# Semantic Specification S — Attribute-Based Authorization Engine (v0)

S = (R, C, I, T, O)

## Purpose

Given a request containing a subject, resource, action, and context, determine
a deterministic authorization decision. The engine is intentionally bounded: it
implements a fixed policy table and does not attempt to be a general-purpose
policy language.

The system is fail closed. It must return an explicit error for malformed or
unknown identifiers; it must return `DENY` when a well-formed request has no
applicable allow rule.

## R — Requirements

1. Given a request, validate that `subject`, `resource`, `action`, and
   `context` have the required JSON shape and accepted values.
2. Reject an unknown subject, resource, action, or malformed context with an
   explicit error. The engine must never convert an invalid request into
   `ALLOW` or an implicit `DENY`.
3. Evaluate every policy rule that matches the request subject, resource,
   action, and all stated context conditions.
4. If one or more matching `DENY` rules exist, return `DENY`, regardless of
   any matching `ALLOW` rules.
5. Otherwise, if one or more matching `ALLOW` rules exist, return `ALLOW`.
6. Otherwise, return `DENY` (default deny).
7. Evaluation must be independent of the ordering of the policy rules.
8. The result must depend only on the request and the fixed policy table; it
   must not depend on wall-clock time, randomness, process state, or previous
   requests.

## C — Contracts / Invariants

- C1: The result is exactly one of `ALLOW`, `DENY`, or an explicit error object.
- C2: For identical valid requests, the result is deterministic.
- C3: No request containing an unknown identifier can return `ALLOW`.
- C4: A matching explicit `DENY` rule overrides every matching `ALLOW` rule.
- C5: A valid request with no matching `ALLOW` rule returns `DENY`.
- C6: Adding, removing, or reordering a policy rule that does not match a
  request cannot change the result for that request.
- C7: Reordering the fixed policy table cannot change the result for any
  request.
- C8: A context-sensitive rule matches only when every stated context condition
  is satisfied.

## I — Interface

Wire format: one JSON request on stdin and one JSON response on stdout.

### Request

```json
{
  "subject": "alice",
  "resource": "report:quarterly",
  "action": "read",
  "context": {
    "network": "corp",
    "hour": 10
  }
}
```

The request fields are:

- `subject`: one of `alice`, `bob`, `carol`, `dave`
- `resource`: one of `report:quarterly`, `report:payroll`, `system:settings`
- `action`: one of `read`, `write`, `approve`
- `context.network`: one of `corp`, `vpn`, `public`
- `context.hour`: integer in the inclusive range 0--23

A missing field, an extra field, an unknown string identifier, a non-integer
hour, or an hour outside 0--23 is malformed or unknown and must return an
explicit error.

### Response

For a valid request:

```json
{"decision": "ALLOW"}
```

or

```json
{"decision": "DENY"}
```

For a malformed request or unknown identifier:

```json
{"error": "Invalid request"}
```

The exact error response above is required for every invalid or unknown input.
The process must exit with a nonzero status when it emits this error response.
It must exit with status zero for `ALLOW` and `DENY` responses.

## Policy Table (v0 — fixed for S0)

Each rule contains an effect and matching conditions. A rule matches only if
all listed conditions match the request. Rules not listed here do not exist.

| Rule | Effect | Subject | Resource | Action | Network | Hour condition |
|---|---|---|---|---|---|---|
| P1 | ALLOW | alice | report:quarterly | read | corp or vpn | 08--18 inclusive |
| P2 | ALLOW | alice | report:quarterly | write | corp | 08--18 inclusive |
| P3 | DENY | alice | report:payroll | read | any | any valid hour |
| P4 | ALLOW | bob | report:quarterly | read | corp | any valid hour |
| P5 | DENY | bob | report:quarterly | write | any | any valid hour |
| P6 | ALLOW | carol | report:payroll | read | corp | 09--17 inclusive |
| P7 | DENY | carol | report:payroll | read | public | any valid hour |
| P8 | ALLOW | dave | system:settings | approve | corp | 08--18 inclusive |
| P9 | DENY | dave | system:settings | approve | any | any valid hour |

The simultaneous presence of P8 and P9 is intentional. For every request to
approve `system:settings` by `dave`, P9 overrides P8 under requirement 4 and
contract C4; the result is therefore always `DENY`.

## T — Validation Requirements

The external validation oracle must be implemented independently of every
generated authorization artifact. It must include the following categories.

### Golden and boundary cases

The golden suite must include at least these cases:

1. `alice` reads `report:quarterly` from `corp` at hour 8: `ALLOW`.
2. `alice` reads `report:quarterly` from `vpn` at hour 18: `ALLOW`.
3. `alice` reads `report:quarterly` from `public` at hour 10: `DENY`.
4. `alice` writes `report:quarterly` from `corp` at hour 18: `ALLOW`.
5. `alice` writes `report:quarterly` from `corp` at hour 19: `DENY`.
6. `alice` reads `report:payroll` from `corp` at hour 10: `DENY`.
7. `bob` reads `report:quarterly` from `corp` at hour 0: `ALLOW`.
8. `bob` reads `report:quarterly` from `vpn` at hour 12: `DENY`.
9. `bob` writes `report:quarterly` from `corp` at hour 12: `DENY`.
10. `carol` reads `report:payroll` from `corp` at hour 9: `ALLOW`.
11. `carol` reads `report:payroll` from `public` at hour 10: `DENY`.
12. `carol` reads `report:payroll` from `corp` at hour 18: `DENY`.
13. `dave` approves `system:settings` from `corp` at hour 10: `DENY`.
14. A valid request with no matching rule returns `DENY`.

### Invalid and unknown-input cases

The oracle must test at least:

- an unknown subject;
- an unknown resource;
- an unknown action;
- an unknown network;
- a missing `context` field;
- a non-integer hour;
- an hour below 0 or above 23;
- an extra top-level request field.

Each case must produce exactly `{"error": "Invalid request"}` and a nonzero
process exit status.

### Property-based tests

The oracle must generate valid and invalid requests and test C1--C8. At a
minimum, it must check determinism, default deny, deny-overrides-allow,
rejection of invalid identifiers, and the non-effect of inapplicable rules.

### Randomized oracle cases

The oracle must execute at least 1,000 seeded randomized requests, including
both valid and invalid requests. Expected results must be computed by an
independently written reference oracle that directly implements this
specification.

## O — Operational Objectives / Preferences

- O1: The engine must process one request in less than 1 ms at the 99th
  percentile on the reference hardware used for the experiment.
- O2: The engine must not access the network, filesystem, clock, environment,
  or external services while evaluating a request.
- O3: The engine must produce exactly one JSON response per input request and
  must not emit diagnostic text on stdout.
- O4: The generated artifact must compile with the documented LLVM toolchain
  and execute as a standalone command-line program.
\end{lstlisting}

\section{Prompt Record for \texttt{authorization\_demo}}
\label{app:authorization-prompts}

This appendix reproduces the prompt used for every independent LLVM generation
in the \texttt{authorization\_demo} experiment. The prompt establishes the
independence constraint, the macOS/Apple Silicon compilation target, the
required stdin/stdout protocol, and the required response format. Its
\texttt{<PASTE spec/S.md HERE>} placeholder was replaced in each generation
with the exact frozen specification reproduced in
Listing~\ref{lst:authorization-specification}.

\subsection{Independent LLVM Generation}

\begin{lstlisting}[
  style=promptstyle,
  caption={Prompt used for each independent LLVM generation of the authorization engine.},
  label={lst:authorization-llvm-generation}
]
# Independent LLVM Generation Prompt — macOS / Apple Silicon

You are a code-generation agent participating in an independent-generation
experiment.

You will be given a semantic specification `S = (R, C, I, T, O)` for a software
component. In the execution protocol, the exact versioned contents of
`spec/S.md` are pasted immediately after this prompt.

## Independence and authority

The supplied `spec/S.md` is the sole authoritative source of semantic
requirements. Generate the implementation directly from that specification.

Do not inspect, translate, imitate, derive from, or use as context any Python,
C, C++, Rust, JavaScript, LLVM, or other implementation of this component. Do
not use an earlier generated artifact. Do not modify the supplied specification,
the policy table, or its validation requirements.

## Required artifact

Generate a complete, self-contained LLVM IR file named
`authorization_vN.ll`, where `N` is the independently assigned generation
number. The file must implement the stdin/stdout JSON protocol defined in the
supplied specification and must be usable as a standalone command-line
executable.

Your response must contain exactly two logical sections:

1. `ASSUMPTIONS`, containing only necessary assumptions caused by genuine
   ambiguity in `spec/S.md`; write `None.` if there are none.
2. `LLVM IR`, containing one fenced code block labelled `llvm` with the
   complete `.ll` file.

Do not provide Python, Rust, C, C++, pseudocode, or a prose implementation.

## Execution environment

The artifact will be compiled and executed locally on this machine:

- Operating system: macOS (Darwin)
- Architecture: Apple Silicon arm64
- Compiler: Apple clang 21.0.0
- LLVM target: `arm64-apple-darwin25.5.0`
- Compilation command:

```sh
clang -O0 -Wno-override-module authorization_vN.ll -o authorization_vN
```

- Execution model: the program receives one JSON request followed by a newline
  on standard input and writes exactly one JSON response followed by a newline
  on standard output.

Do not assume Linux, GNU libc, ELF, `/proc`, Linux system calls, or a Linux
linker. This is a Darwin Mach-O target, not a Linux/ELF target. Emit these
module declarations exactly at the start of the generated file:

```llvm
target datalayout = "e-m:o-p270:32:32-p271:32:32-p272:64:64-i64:64-i128:128-n32:64-S128-Fn32"
target triple = "arm64-apple-macosx26.0.0"
```

For line-oriented JSON input, use the Darwin C runtime symbol `@__stdinp`, not
`@stdin`: on macOS, `stdin` is exposed through the `__stdinp` data symbol. Use
this exact declaration when calling `fgets`:

```llvm
@__stdinp = external global ptr, align 8
declare ptr @fgets(ptr nocapture noundef, i32 noundef, ptr nocapture noundef) nounwind
```

Load the stream pointer before passing it to `fgets`, for example:

```llvm
%stream = load ptr, ptr @__stdinp, align 8
%result = call ptr @fgets(ptr %buffer, i32 4096, ptr %stream)
```

Do **not** declare or reference `@stdin`, `@stdout`, `@stderr`, `stdin`,
`stdout`, or `stderr`; those Linux-style symbol names will fail to link on this
Darwin target. Do not use glibc-specific functions or assumptions. Standard
C-library functions such as `fgets`, `strcmp`, `strstr`, `strtol`, `printf`,
and `puts` may be declared with ABI-compatible LLVM IR signatures. Do not use
target-specific assembly or external libraries.

## Implementation constraints

- Implement the JSON acceptance and rejection behavior required by `spec/S.md`.
- Reject malformed input, unknown identifiers, missing fields, extra top-level
  fields, non-integer hours, and out-of-range hours exactly as specified.
- Emit exactly `{"error": "Invalid request"}` and exit nonzero for invalid
  requests.
- Emit exactly one of `{"decision": "ALLOW"}` or `{"decision": "DENY"}`
  and exit zero for valid requests.
- Enforce deny-overrides-allow and default-deny semantics exactly.
- Do not depend on input field ordering unless `spec/S.md` explicitly permits
  it; accept the JSON request shape specified by the interface.
- Do not access the network, filesystem, clock, environment, or external
  services during request evaluation.
- Keep output deterministic and do not print diagnostics to stdout.
- Produce actual valid LLVM IR; do not embed C, C++, Rust, Python, or
  pseudocode in the file.

## Required internal self-check

Before producing the response, verify internally that the LLVM IR:

1. compiles with the stated Apple clang command on macOS arm64;
2. implements every requirement and contract in the supplied `spec/S.md`;
3. treats the policy table as unordered in meaning;
4. returns `DENY` for the intentional P8/P9 conflict;
5. uses no prior implementation as an intermediate representation; and
6. contains a complete standalone implementation.

## Semantic specification

Paste the exact, unmodified contents of `spec/S.md` below this line:

<PASTE spec/S.md HERE>
\end{lstlisting}

\end{document}